\documentstyle[12pt,epsf]{article}

\begin{document}


\newcommand{\be}{\begin{equation}}
\newcommand{\ee}{\end{equation}}
\newcommand{\<}{\langle}
\renewcommand{\>}{\rangle}
\newcommand{\reff}[1]{(\ref{#1})}

\def\bsigma{\mbox{\protect\boldmath $\sigma$}}

\def\spose#1{\hbox to 0pt{#1\hss}}
\def\ltapprox{\mathrel{\spose{\lower 3pt\hbox{$\mathchar"218$}}
 \raise 2.0pt\hbox{$\mathchar"13C$}}}
\def\gtapprox{\mathrel{\spose{\lower 3pt\hbox{$\mathchar"218$}}
 \raise 2.0pt\hbox{$\mathchar"13E$}}}
\def\inapprox{\mathrel{\spose{\lower 3pt\hbox{$\mathchar"218$}}
 \raise 2.0pt\hbox{$\mathchar"232$}}}

\title{ \vspace*{-2.5cm}{\small\rm \noindent IFUP Th-58/92
                         \hfill SNS 16/92}\\[1cm]
      Possible failure of asymptotic freedom in two-dimensional
      $RP^2$ and $RP^3$ $\sigma$-models}

\author{
  \\
  {\small Sergio Caracciolo}              \\[-0.2cm]
  {\small\it Scuola Normale Superiore and INFN -- Sezione di Pisa}  \\[-0.2cm]
  {\small\it Piazza dei Cavalieri}        \\[-0.2cm]
  {\small\it Pisa 56100, ITALIA}          \\[-0.2cm]
  {\small Internet: {\tt CARACCIO@UX1SNS.SNS.IT}}     \\[-0.2cm]
  {\small Bitnet:   {\tt CARACCIO@IPISNSVA.BITNET}}   \\[-0.2cm]
  {\small Hepnet/Decnet:   {\tt 39198::CARACCIOLO}}   \\[-0.2cm]
  \\[-0.1cm]  \and
  {\small Robert G. Edwards}              \\[-0.2cm]
  {\small\it Supercomputer Computations Research Institute}         \\[-0.2cm]
  {\small\it  Florida State University}   \\[-0.2cm]
  {\small\it Tallahassee, FL 32306 USA}   \\[-0.2cm]
  {\small Internet: {\tt EDWARDS@MAILER.SCRI.FSU.EDU}} \\[-0.2cm]
  \\[-0.1cm]  \and
  {\small Andrea Pelissetto}              \\[-0.2cm]
  {\small\it Dipartimento di Fisica}      \\[-0.2cm]
  {\small\it Universit\`a degli Studi di Pisa}        \\[-0.2cm]
  {\small\it Pisa 56100, ITALIA}          \\[-0.2cm]
  {\small Internet: {\tt PELISSET@SUNTHPI1.DIFI.UNIPI.IT}}   \\[-0.2cm]
  {\small Bitnet:   {\tt PELISSET@IPISNSVA.BITNET}}   \\[-0.2cm]
  {\small Hepnet/Decnet:   {\tt 39198::PELISSETTO}}   \\[-0.2cm]
  {\protect\makebox[5in]{\quad}}  
  \\[-0.1cm]  \and
  {\small Alan D. Sokal}                  \\[-0.2cm]
  {\small\it Department of Physics}       \\[-0.2cm]
  {\small\it New York University}         \\[-0.2cm]
  {\small\it 4 Washington Place}          \\[-0.2cm]
  {\small\it New York, NY 10003 USA}      \\[-0.2cm]
  {\small Internet: {\tt SOKAL@ACF3.NYU.EDU}}        \\[-0.2cm]
  {\protect\makebox[5in]{\quad}}  
  \\[0.5cm]
{\small Talk given at the Lattice '92 Conference held in Amsterdam}
}

\date{}

\maketitle
\thispagestyle{empty}   

\vspace*{2cm}
\begin{abstract}
We have simulated the two-dimensional $RP^2$ and $RP^3$ $\sigma$-models,
at correlation lengths up to about 220 (resp.\ 30), using a Wolff-type
embedding algorithm.  We see no evidence of asymptotic scaling.
Indeed, the data rule out the conventional asymptotic scaling scenario
at all correlation lengths less than about $10^{9}$ (resp.\ $10^{5}$).
Moreover, they are consistent with a critical point at $\beta \approx 5.70$
(resp.\ 6.96), only 2\% (resp.\ 5\%) beyond the largest $\beta$
at which we ran.  Preliminary studies of a mixed $S^{N-1}/RP^{N-1}$ model
(i.e. isovector + isotensor action) show a similar behavior when
$\beta_T \to \infty$ with $\beta_V$ fixed $\ltapprox 0.6$,
while they are consistent with
conventional asymptotic freedom along the lines
$\beta_T/\beta_V$ fixed $\ltapprox 2$.  Taken as a whole,
the data cast doubt on (though they do not completely exclude)
the idea that $RP^{N-1}$ and $S^{N-1}$ $\sigma$-models lie in the same
universality class.
\end{abstract}

\clearpage


The present work grows out of our study of Wolff-type embedding
algorithms for general nonlinear $\sigma$-models
\cite{Edwards_89,CEPS_swwo4c2}.
Our main result
\cite{CEPS_swwo4c2}
was that a Wolff-type embedding algorithm
can perform well --- in the sense of having a dynamic critical exponent
$z \ll 2$ --- only if it is based on an involutive isometry
of the target manifold $M$ whose fixed-point manifold has codimension 1.
Such an involutive isometry exists
only if $M$ is a sphere, a real projective space,
or a discrete quotient of products of such spaces.

After extensive studies of the models where $M$ is a sphere $S^{N-1}$
(the so-called $N$-vector model), we considered next the models with
$M=RP^{N-1}=S^{N-1}/Z_2$ \cite{CEPS_LAT91}. We were thus led to re-examine the
long-standing puzzle concerning the phase diagram of these models.

This interest is not purely academic.
$RP^2$
$\sigma$-models are used as
models for nematic liquid crystals, where instead of spins one deals with the
orientation of elongated molecules
exhibiting a head-tail ($\bsigma \to -\bsigma$) symmetry.

On a $Z^2$ lattice we shall consider (for the moment)
\be
- \beta  H_T (\{\bsigma\})   \;=\;
    {\beta \over 2} \sum_{\< xy \>} (\bsigma_x \cdot \bsigma_y)^2
 \label{ham:tensor}
\ee
(T for ``tensor'') with $\bsigma_x\in S^{N-1}$.
Clearly $H_T$
is invariant under the $Z_2$ gauge  transformations
\be
\bsigma_x \to \eta_x \bsigma_x
\ee
with $\eta_x \in Z_2 = \{\pm1\}$.
It follows that all non-gauge-invariant correlation functions vanish
identically, for example
\be
\< \sigma_x^a \sigma_y^b \> = 0 \qquad \hbox{if } x \neq y   \;.
\ee
The fundamental correlation function is thus the isotensor
\be
G_T(x,y) = \< (\bsigma_x \cdot \bsigma_y)^2 \> - 1/N   \;.
\ee

In the {\em formal}\/ continuum limit,
the discrete gauge symmetry plays no role.
Locally the spaces $S^{N-1}$ and $RP^{N-1}$ are the same,
and one thus expects the model \reff{ham:tensor}
to have the same continuum limit {\em in the isotensor sector}\/
as the ordinary $N$-vector model.

At $N=\infty$, the model \reff{ham:tensor} undergoes a first-order transition
\cite{DiVecchia_84};
but this transition involves the spontaneous breaking of the
$Z_2$ gauge symmetry, and thus cannot persist to finite $N$.
For finite $N$, some workers have found
(by high-temperature expansions or Monte Carlo simulations)
indications of a
second-order phase transition, which could be related~\cite{Kunz_91} to a
condensation of $Z_2$-vortices in the field
\be
A_\mu(x) = \hbox{sign }  [ \bsigma_x \cdot \bsigma_{x+\mu} ]   \;.
\ee
If there is not a phase transition, the model is presumably
{\em asymptotically free}\/, and a renormalization-group calculation
predicts that for $N > 2$
the correlation length $\xi$ and susceptibility $\chi$ behave as
\begin{eqnarray}
\xi^{af}(\beta)    & \sim &   C_\xi  \beta^{-1/(N-2)} e^{2\pi\beta/(N-2)}
        \label{O(N)_xi_predicted}  \\
\chi^{af}(\beta)   & \sim &   C_\chi  \beta^{-(N+1)/(N-2)} e^{4\pi\beta/(N-2)}
        \label{O(N)_chi_predicted}
\end{eqnarray}
up to corrections of order $1/\beta$.

By using the exact results for the mass gap for the $O(N)$-models
\cite{Hasenfratz-Niedermayer} and the ratios of
$\Lambda$-parameters in different schemes (which are extracted from first-order
perturbation theory), it is even possible to predict what is the value of the
constant $C_\xi$ for the exponential correlation length (inverse mass gap).
In the following we shall make use of the so-called second-moment correlation
length,
which should differ by a few percent from the exponential correlation length.

In Fig.~\ref{rp_af} we report the ratio of the measured correlation length
to the value~\reff{O(N)_xi_predicted} predicted by asymptotic freedom,
as a function of $\beta$ on lattices of sizes $32\leq L\leq 512$,
for both
$RP^2$ and $RP^3$. The observed correlation length is smaller than the expected
value by a factor of about $10^7$ \cite{Sinclair_82} (resp. $10^4$).
The simulations at the larger values of $\beta$ and $L$ clearly show that the
asymptotic regime has not been reached, as the curves show a clear rise up.
In any case,
our simulations on {\em small}\/ lattices prove that
{\em if}\/ the conventional scenario is true,
the asymptotic regime cannot be seen on lattices of size less than about
$10^{9}$  (resp. $10^{5}$) --- a result which is quite astonishing.
\begin{figure}[p]
\vspace*{0cm} \hspace*{-0cm}
\epsfxsize= \hsize
\caption{
Asymptotic freedom test for $RP^2$ and $RP^3$.
 } \label{rp_af}
\end{figure}

Let us assume, instead, that there is a phase transition
at a finite $\beta=\beta_c$.
Then we expect by finite-size scaling that
\begin{eqnarray}
\xi(\beta)  & = &  L f_\xi  \!\left( |\beta-\beta_c|^\nu L \right)
        \label{FSS_xi_predicted} \\
\chi(\beta) & = &  L^{\gamma/\nu} f_\chi \!\left( |\beta-\beta_c|^\nu L \right)
        \label{FSS_chi_predicted}
\end{eqnarray}
where $L$ is the size of the lattice, and $\nu$ and $\gamma$ are, as usual, the
critical exponents for the correlation length and the susceptibility.
Thus we could try to fix $\beta_c$ and $\nu$
by using the data for $\xi$ on different
lattice sizes and by requiring that all the obtained values of $\xi/L$,
when plotted
as a function of $(1-\beta/\beta_c) L^{1/\nu}$, fall on the same curve.
Analogously, once $\beta_c$ and $\nu$ have been fixed, $\gamma$ can be found
by imposing \reff{FSS_chi_predicted}.

In Fig.~\ref{csi_scaling} we have plotted the observed values of $\xi/L$
as a function of $(1-\beta/\beta_c) L^{1/\nu}$ with $\beta_c=5.70$ and
$\nu=2.0$  (see also~\cite{Kunz_91}).  The same analysis for $\chi$ fixes
$\gamma\approx 3.3$.
\begin{figure}[p]
\vspace*{0cm} \hspace*{-0cm}
\epsfxsize= \hsize
\caption{
Scaling for the correlation length for $RP^2$.
 } \label{csi_scaling}
\end{figure}
Also for $RP^3$ it is possible to fit the data by supposing a $\beta_c=6.96$
and the same values of $\nu$ and $\gamma$ as for $RP^2$.

The quite large values of the critical exponents may suggest
a phase transition  of the Kosterlitz-Thouless type: in this case
the behaviour given in \reff{FSS_xi_predicted} and \reff{FSS_chi_predicted}
should be
substituted by
\be
\xi(\beta) \, = \,
L f_\xi  \!\left( \exp\!\left[ -{A (1-\beta /\beta_c)^{-1/2}} \right] L \right)
        \label{KT_xi_predicted}
\ee
\be
\chi(\beta)   \, = \,
L^{\gamma/\nu}
f_\chi \!\left( \exp\!\left[- {A (1-\beta /\beta_c)^{-1/2}} \right] L \right)
        \label{KT_chi_predicted}
\ee
where \reff{KT_xi_predicted} should be used to determine the constants $A$ and
$\beta_c$.
In Fig.~\ref{csi_scaling_KT} we have plotted the observed values of $\xi/L$
as a function of $\exp ( -{A/ |1-\beta /\beta_c|^{1/2}}) L$ with
$\beta_c=5.85$ and $A=1.73$. The same analysis for $\chi$ fixes
$\gamma/\nu\approx 1.7$.
\begin{figure}[p]
\vspace*{0cm} \hspace*{-0cm}
\epsfxsize= \hsize
\caption{
Scaling of the Kosterlitz-Thouless type for the correlation length for $RP^2$.
 } \label{csi_scaling_KT}
\end{figure}
Also for $RP^3$ it is possible to fit the data by supposing a $\beta_c=7.15$
and the  values  $A=1.7$ and $\gamma/\nu=1.65$.

A plot of $\chi/L^{\gamma/\nu}$ as a function of $\xi/L$ provides directly
an estimate of $\gamma/\nu$, which is independent of the assumptions
on the scaling behaviour. See Fig.~\ref{gammasunu_rp2}.
\begin{figure}[p]
\vspace*{0cm} \hspace*{-0cm}
\epsfxsize= \hsize
\caption{
$\chi/L^{1.7}$ vs. $\xi/L$ for $N=3$ with $0\leq \beta_V \leq 0.5$. }
\label{gammasunu_rp2}
\end{figure}

Next we considered the Hamiltonian
\begin{eqnarray}
\lefteqn{  - \beta H_{mixed} (\{\bsigma\})   \;=\;}\nonumber \\
& & =   \beta_V \sum_{\< xy \>} \bsigma_x \cdot \bsigma_y
    \;
    +\, {\beta_T \over 2} \sum_{\< xy \>} (\bsigma_x \cdot \bsigma_y)^2
 \label{ham:mixed}
\end{eqnarray}
(V for ``vector") which interpolates between the models with target space
$S^{N-1}$ and $RP^{N-1}$.
The normalization of the two terms has been chosen in order to get
$\beta=\beta_V+\beta_T$ as the coupling constant in the
usual low-temperature expansion.

For $N=4$, we made simulations
along the lines with  $r = \beta_T/\beta_V$ fixed and not too large
(0.5, 1.0, 2.0).
We found that the model behaves
qualitatively in agreement with the conventional scenario of asymptotic freedom
in the sense that
\begin{trivlist}
\item[$\bullet$] the ratio $\xi/\xi^{af}$ is not
very far from 1, although it becomes worse and worse when $r$ increases and
at $r=2.0$ it is around 0.4;
\item[$\bullet$] the estimated values of $\beta_c$ which appear when we try the
Ans\"atze \reff{FSS_xi_predicted}/\reff{FSS_chi_predicted}
 or \reff{KT_xi_predicted}/\reff{KT_chi_predicted}
are relatively far from the actual values of
$\beta$ which can be simulated effectively.
But this gap decreases as $r$ increases;
\item[$\bullet$] the ratio of the vectorial
over the tensorial correlation length seems to approach a constant value
in the continuum limit;
\item[$\bullet$] the dynamic behaviour of the Wolff embedding algorithm
is in the same universality class as for the $O(N)$ model,
which means that the dynamic critical
exponent is almost zero (critical slowing-down is eliminated).
\end{trivlist}

\begin{figure}[p]
\vspace*{0cm} \hspace*{1cm}
\epsfxsize= \hsize
\vspace*{-0.9cm}\caption{
A possible phase diagram. }
\label{phase_diagram}
\end{figure}

On the other hand,
in the region where $\beta_T \to \infty$ with $\beta_V$ fixed and small
($\ltapprox 0.6$), the model looks
very similar to the $RP^{N-1}$ case: the ratio $\xi/\xi^{af}$ is
extremely small, and the estimated values of a transition point $\beta_c$
are very close to the largest values of $\beta$ at which we ran.
The vectorial sector
appears decoupled from the tensorial one: the vectorial correlation length and
susceptibility appear to approach finite values
even when the corresponding tensorial
quantities start to diverge. Also from the dynamical  point of view the
behaviour is the same as in the $RP^{N-1}$ case,
with a dynamic critical exponent
for the tensorial susceptibility close to 1~\cite{CEPS_LAT91}.

But we have also seen, in our Monte Carlo runs for $N=3$,
that there is a crossover,
near $\beta_V \approx 0.5$,
where also the vectorial sector begins to become critical.
Ising-like fluctuations seem to become important in this regime, so that the
Wolff embedding algorithm is less efficient.
We are currently investigating this region:
it would be interesting to understand
whether the apparent transition curve which starts on the axis
$\beta_V=0$ stops somewhere,
and whether the Ising critical point which occurs at
($\beta_T=\infty$, $\beta_V= 0.44\ldots$) is isolated
or belongs to a phase-separation curve extending to finite $\beta_T$.
The phase diagram could appear like in Fig.~\ref{phase_diagram},
but are the phase transitions actually there?
An unconventional scenario has been suggested~\cite{Seiler}
in which asymptotic freedom
does not hold even in the pure vector model. Indeed, conjectured
correlation inequalities,
based on the remark that the couplings are both ferromagnetic,
would suggest that
 the line where the tensorial correlation length diverges ---
        if there is one --- cannot stop
somewhere in the
middle of the phase diagram,
and must at least intersect every ray
with $r = \beta_T/\beta_V$ fixed $>0$.
We are continuing to run our simulation,
and hope to add other pieces of evidence to
clarify these important open questions.

\section*{Acknowledgments}

The authors' research was supported in part by
U.S.\ National Science Foundation grants DMS-8911273 and DMS-9200719,
U.S.\ Department of Energy contract DE-FG02-90ER40581  and
DE-FC05-85ER250000,
NATO Collaborative Research Grant CRG 910251.

\end{document}